\documentstyle[preprint,aps, epsfig]{revtex}

\begin{document}


\preprint{
\hfill$\vcenter{\hbox{\bf IFUSP/P-1315} 
                \hbox{\bf IFT-P.054/98} 
                \hbox{\bf hep-ph/9808288}
             }$ }

\title{Multilepton Signatures for Leptoquarks}

\author{O.\ J.\ P.\ \'Eboli$^1$\thanks{Email: eboli@ift.unesp.br}, 
R.\ Z.\ Funchal$^2$\thanks{Email: zukanov@charme.if.usp.br}, 
and T.\ L.\   Lungov$^1$\thanks{Email: thais@ift.unesp.br} }

\address{\em $^1$ Instituto de F\' {\i}sica Te\'orica -- UNESP \\
    R. Pamplona 145, 01405-900 S\~ao Paulo, Brazil \\[8pt]
    $^2$ Instituto de F\'{\i}sica, Universidade de S\~ao Paulo, \\
    C.\ P.\ 66.318, 05389-970 S\~ao Paulo, Brazil. }

\maketitle

\vspace{.2in}


\hfuzz=25pt

\begin{abstract}

 The production of third generation leptoquarks can give rise to
 multilepton events accompanied by jets and missing $E_T$. In this
 work we study the signals of these leptoquarks at the CERN Large
 Hadron Collider and compare them with the ones expected in
 supersymmetric models.

\end{abstract}

\newpage

\section{Introduction}

Many theories, like composite models~\cite{comp,af}, technicolor~\cite{tec}, 
and grand unified theories~\cite{gut}, predict the
existence of new particles, called leptoquarks, that mediate
quark-lepton transitions. In this work we focus our attention to
scalar leptoquarks ($S$) that couple to pairs $t$--$\ell$ or
$b$--$\ell$ with $\ell = e$, $\mu$ or $\tau$.  At
the CERN Large Hadron Collider (LHC), leptoquarks can be pair produced
by gluon--gluon and quark--quark fusions, as well as singly produced
in association with a lepton in gluon--quark reactions. Therefore, the
production of third generation leptoquarks can lead to multilepton
signals accompanied by jets and missing $E_T$ ($\not{\!\!E_T}$) since
the heavy quark decay can give rise to further leptons and jets.  This
means that third generation leptoquarks can, in principle, mimic the
multilepton supersymmetry (SUSY) signals~\cite{paige}.  For this reason, we
investigated the importance of the multilepton signatures for such
leptoquarks at the LHC.

In our analyses we considered the following multilepton topologies:

$\bullet$ One lepton topology (1L) which exhibits one lepton ($e^\pm$
or $\mu^\pm$) in association with jets and $\not{\!\!E_T}$;

$\bullet$ opposite-sign dilepton events (OS) which contain a pair
of leptons of opposite charge in addition to jets and $\not{\!\!E_T}$;

$\bullet$  same-sign  dilepton topology (SS) which presents a pair
of leptons with the same charge, jets and $\not{\!\!E_T}$;

$\bullet$ trilepton events (3L) which possess 3 charged leptons, jets,
and $\not{\!\!E_T}$.

\noindent Moreover, we employed the cuts of Ref.\ \cite{paige} 
which studied the multilepton signals for supersymmetry in the
framework of the minimal supergravity model (mSUGRA). The use of these
cuts not only reduces the standard model (SM) backgrounds, but also
allow us to compare the leptoquark signals with the mSUGRA ones.

In principle, leptoquark events possess the striking signature of a
peak in the invariant mass of a charged lepton and a jet, which could
be used to further reduce backgrounds and to establish that an
observed signal is due to leptoquarks. This is an important feature of
the signals for first generation leptoquarks~\cite{fut:pp}.
Notwithstanding, third generation leptoquarks exhibit cascade decays
containing heavy quarks and/or $\tau^\pm$, which give rise to
neutrinos, and consequently wash out the lepton-jet invariant mass
peak.

Since leptoquarks are an undeniable signal of physics beyond the SM, there
have been several direct searches for them in accelerators. At the Tevatron
collider it was established that leptoquarks coupling to $b$--$\tau$ pairs
should be heavier than 99~GeV~\cite{teva}.  Moreover, low-energy experiments
lead to indirect bounds on the couplings and masses of third generation
leptoquarks.  Leptoquarks may give rise to flavor changing neutral current
processes if they couple to more than one family of quarks or
leptons~\cite{shanker,fcnc}.  In order to avoid these bounds, we assumed that
the leptoquarks couple only to one quark family and one lepton generation. The
effects of third generation leptoquarks on the $Z$ physics through radiative
corrections lead to limits on leptoquarks that couple to top
quarks~\cite{gbjkm}.  As a rule of a thumb, the $Z$-pole data constrain the
masses of leptoquarks to be larger than $200$---$500$ GeV when their Yukawa
coupling is equal to the electromagnetic coupling $e$~\cite{gbjkm,leurer}.

\section{Analyses}

A natural hypothesis for theories beyond the SM is that they exhibit
the gauge symmetry $SU(2)_L \otimes U(1)_Y$ above the electroweak
symmetry breaking scale $v$, therefore, we imposed this symmetry on
the leptoquark interactions.  In order to evade strong bounds coming
from the proton lifetime experiments, we required baryon ($B$) and
lepton ($L$) number conservation.  The most general effective
Lagrangian for leptoquarks satisfying the above requirements and
electric charge and color conservation is given by~\cite{buch}
\begin{eqnarray}
{\cal L}_{{eff}}~  &=& {\cal L}_{F=2} ~+~ {\cal L}_{F=0} 
\; , 
\label{e:int}
\\
{\cal L}_{F=2}~  &=& g_{{1L}}~ \bar{q}^c_L~ i \tau_2~ 
\ell_L ~S_{1L}+ 
g_{{1R}}~ \bar{u}^c_R~ e_R ~ S_{1R} 
+ \tilde{g}_{{1R}}~ \bar{d}^c_R ~ e_R ~ \tilde{S}_1
\nonumber \\
&& +~ g_{3L}~ \bar{q}^c_L~ i \tau_2~\vec{\tau}~ \ell_L \cdot \vec{S}_3 
\; ,
\nonumber \\
{\cal L}_{F=0}~  &=& h_{{2L}}~ R_{2L}^T~ \bar{u}_R~ i \tau_2 ~
 \ell_L 
+ h_{{2R}}~ \bar{q}_L  ~ e_R ~  R_{2R} 
+ \tilde{h}_{{2L}}~ \tilde{R}^T_2~ \bar{d}_R~ i \tau_2~ \ell_L
\nonumber
\end{eqnarray}
where $F=3B+L$, $q$ ($\ell$) stands for the left-handed quark (lepton)
doublet, and we omitted the flavor indices of the leptoquark couplings
to fermions. The leptoquarks $S_{1R(L)}$ and $\tilde{S}_1$ are
singlets under $SU(2)_L$, while $R_{2R(L)}$ and $\tilde{R}_2$ are
doublets, and $S_3$ is a triplet.

The multilepton samples due to leptoquarks were obtained using the
Monte Carlo event generator PYTHIA~\cite{pythia}. We assumed in our
analyses  that the leptoquarks decay exclusively into a
single quark-lepton pair. The general case can be easily obtained by
multiplying the signal cross section by an appropriate branching
ratio, which can be read from the lagrangian~(\ref{e:int}).

The cross sections for leptoquark ($S_{{lq}}$) pair production via $q
+\bar{q} \rightarrow S_{{lq}} + \bar{S}_{{lq}}$ or $g + g \rightarrow
S_{{lq}} + \bar{S}_{{lq}}$ are model independent because the
leptoquark--gluon interaction is entirely determined by the $SU(3)_C$
gauge invariance. On the other hand, the single production through $q
+ g \rightarrow S_{{lq}} + \ell$ is model dependent once it involves
the unknown Yukawa coupling of leptoquarks to a lepton--quark
pair. However, this last process is important only for third
generation leptoquarks coupling to $b$ quarks since the top quark
content of the proton is negligible at the LHC energy.

In this work we focused our attention on leptoquarks decaying into
$b$--$\ell$ or $t$--$\ell$ pairs, with $\ell = e$, $\mu$, $\tau$.
We generated samples containing 10\,000 events for each
leptoquark type and production mechanism, assuming two values for the
masses: 300 and 500 GeV. Since the $b$ content of the proton is rather
small, we assumed that the leptoquark Yukawa coupling to be 10 times
de electron electric charge for the single production of $b$--$\tau$,
$b$--$\mu$, or $b$--$e$ leptoquarks.

In our analyses, we applied the following cuts used in Ref.\ \cite{paige}:

\begin{itemize}

\item clusters with  $E_T > 100$~GeV and $|\eta(\mbox{jet})|<3$ are 
labeled as jets; however, for jet-veto only, clusters with
$E_T>25$~GeV and $|\eta(\mbox{jet})|<3$ are regarded as jets;

\item muons and electrons are classified as isolated if they have $p_T >
10$~GeV, $|\eta(l)|<2.5$ and the visible activity within a cone of
$R=\sqrt{\Delta \eta^2+\Delta\Phi^2}=0.3$ about the lepton direction
is less then $E_T(\mbox{cone})=5$~GeV;

\item jet multiplicity, $n_{\mbox{jet}} \geq 2$, with $E_T(\mbox{jet})>
100$~GeV;

\item transverse sphericity $S_T > 0.2$,

\item $E_T(j_1)$, $E_T(j_2) > E_T^c$ and $\not{\!\!E_T}$ $>
E_T^c$, where $E_T^c$ is a parameter that one can vary, see the
figures below;

\item we required the leptons to have $p_T(l)>20$~GeV and
$M_T(l,$$\not{\!\!E_T}$)$>100$~GeV for the one lepton signal and
$p_T(l_{1(2)})$~GeV for $n=2,3$ lepton signals.

\end{itemize}

In our analyses, we simulated a simple calorimeter using the subroutine LUCELL
, which is part of JETSET/PYTHIA package, adopting the same parameters
employed in Ref.\ \cite{paige}.  We should also point out that the effect of
cracks, edges and other detector inefficiencies have not been taken into
account here.
 
\section{Results}

In the following figures we present our results for the leptoquark
cross sections after the above cuts as a function of the parameter
$E_T^c$. For the sake of comparison, we also exhibit in our figures
the SM backgrounds (BG) and mSUGRA cross sections for two sets of
parameters chosen in Ref.\ \cite{paige}, which correspond to the
extreme cases analyzed in this work.  In case 1, it is assumed that
$m_0=m_{\frac{1}{2}}=100$ GeV, $m_{\tilde{g}}=290$ GeV, and
$m_{\tilde{q}}=270$ GeV, while, in case 6, $m_0=4m_{\frac{1}{2}}=2000$
GeV, $m_{\tilde{g}}=1300$ GeV, and $m_{\tilde{q}}=2200$ GeV. Both
scenarios employ $A_0=0$, $\tan \beta=2$, and $m_t=170$ GeV.

The production cross section for $b$--$\mu$ and $b$--$e$ leptoquarks are the
same. Moreover, this is also true for the production of $t$--$e$ and
$t$--$\mu$ leptoquarks. Therefore we present our results only for $b$--$e$ and
$t$--$e$ leptoquarks which are equal to the ones for $b$--$\mu$ and $t$--$\mu$
leptoquarks respectively.

\begin{figure}
\parbox[c]{6.0in}{
\mbox{\qquad\epsfig{file=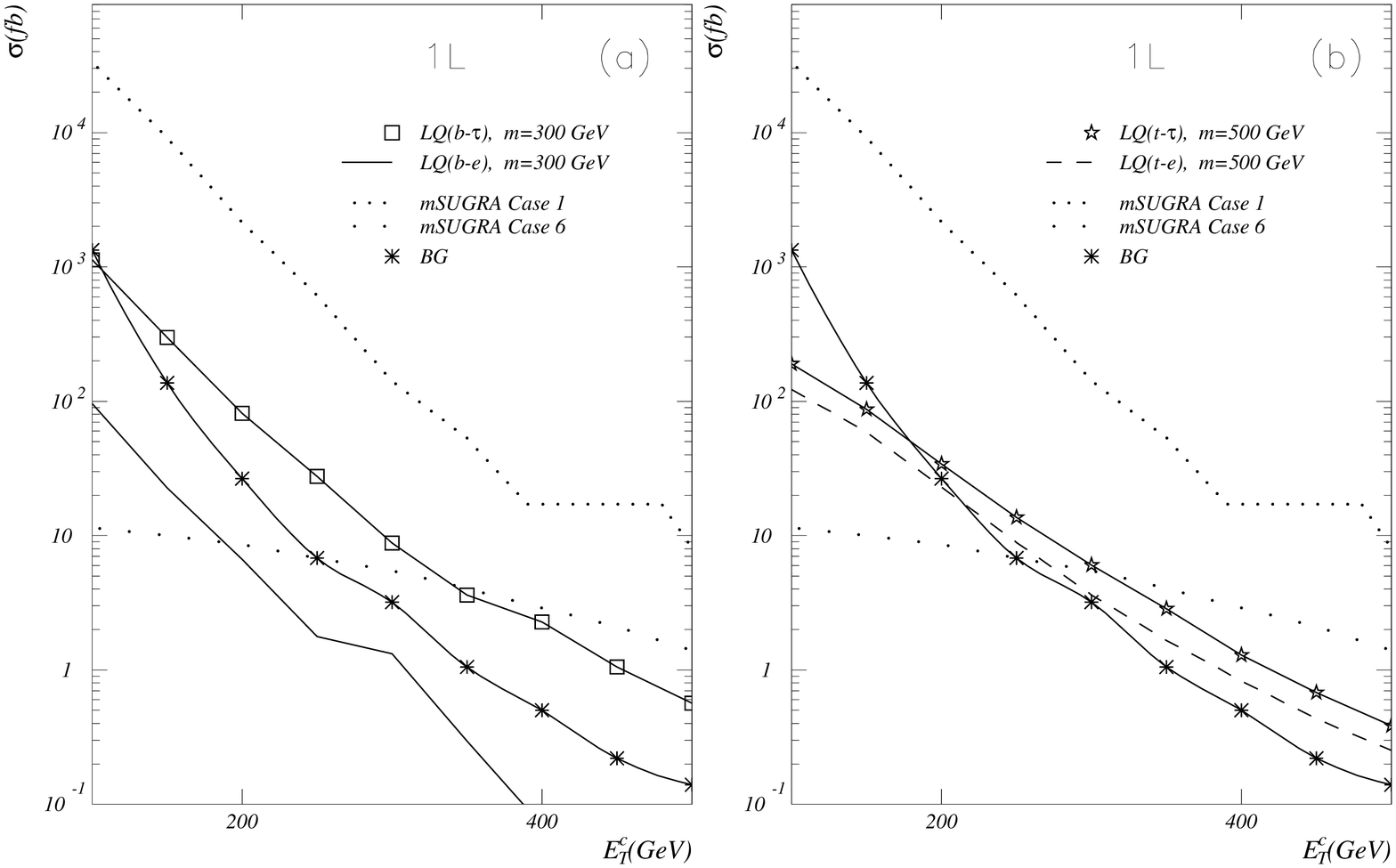,width=\linewidth}}
}
\caption{Production
  cross sections of 1L events as a function of $E_T^c$ for the SM backgrounds
  and two sets of mSUGRA parameters (case 1 and case 6). (a) also contains the
  results for $b$--$e$ and $b$--$\tau$ leptoquarks with a mass of 300~GeV
  while (b) presents the results for $t$--$e$ and $t$--$\tau$ leptoquarks with
  a mass of 500~GeV.}
\label{1ljets}
\end{figure}

We show in Fig.\ \ref{1ljets}(a) the leptoquark production cross sections 
into the 1L topology as a function of $E_T^c$ for scalar leptoquarks
decaying into $b$--$e$ and $b$--$\tau$ with a mass of 300 GeV. 
As we can see, the $b$--$e$ signal is immersed
in the SM backgrounds since the  $E_T$ cuts affects strongly the
signal. On the other hand, the $b$--$\tau$ signal is well
above the background for all values of the parameter
$E_T^c$. Furthermore, the $b$--$\tau$ leptoquarks lead to cross
sections with values between the two mSUGRA cases for $E_T^c \lesssim
350$ GeV. In Fig.\ \ref{1ljets}(b) we present the results for the 1L
topology in the case of $t$--$e$ and $t$--$\tau$ leptoquarks with
masses of 500 GeV. In this case, the signals are always above the 
background and are also between the two mSUGRA cases.

\begin{figure}
\parbox[c]{6.0in}{
\mbox{\qquad\epsfig{file=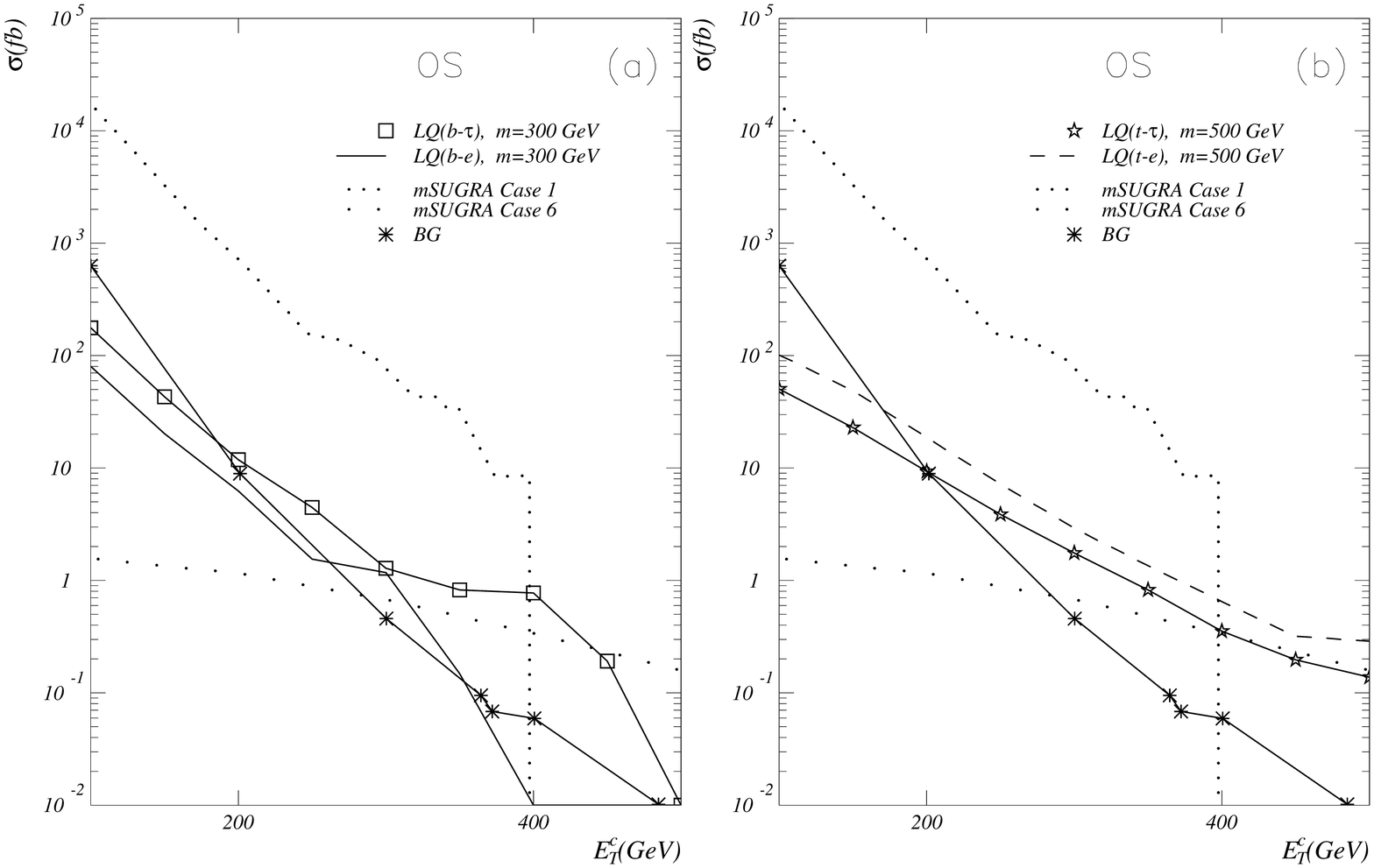,width=\linewidth}}
  }
\caption{Same as Fig.\ \protect\ref{1ljets} for  OS events.}
\label{osjets}
\end{figure}

In Fig.\ \ref{osjets}(a) we present our results for scalar $b$--$e$ and
$b$--$\tau$ leptoquarks with a mass of 300~GeV into the OS topology.  We
observe again that the signal for $b$--$e$ leptoquarks is practically always
immersed in the SM backgrounds due to the same reason pointed out in the
previous case.  The $b$--$\tau$ leptoquark signal is above the background for
$E_T^c > 200$ GeV. This type of leptoquark leads to a cross section with
values between the two mSUGRA cases independently of the $E_T^c$ cut applied.
In Fig.\ \ref{osjets}(b) the results for the same topology but for $t$--$e$
and $t$--$\tau$ leptoquarks with a mass of 500 GeV are displayed.  For $E_T^c
> 200$ GeV these two signals are above the expected BG and their cross section
values are between the two mSUGRA extreme cases even if one imposes a large
$E_T^c$ cut.

\begin{figure}
\parbox[c]{6.0in}{
\mbox{\qquad\epsfig{file=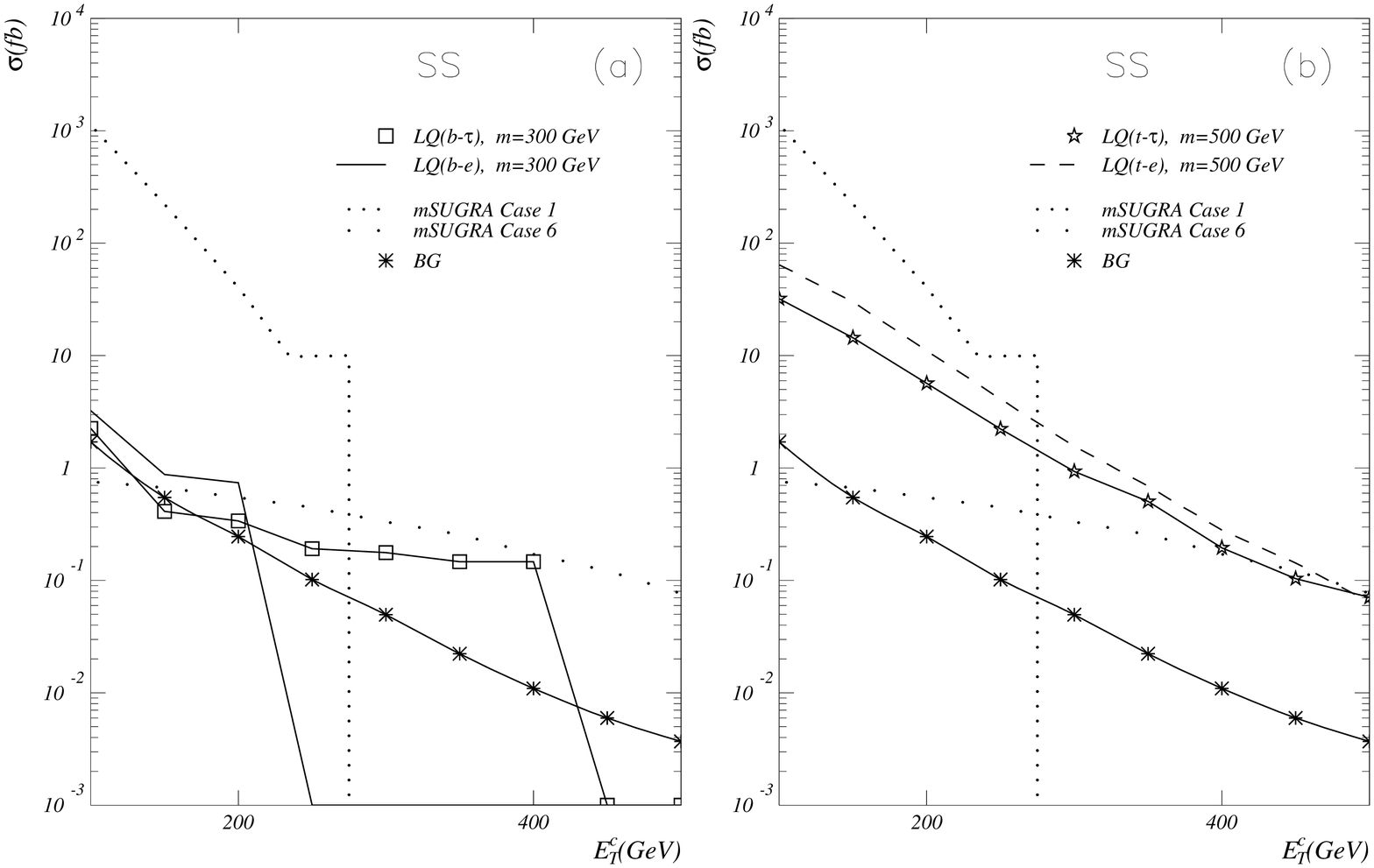,width=\linewidth}}
  }
\caption{Same as Fig.\ \protect\ref{1ljets} for SS events.}
\label{ssjets}
\end{figure}

The production cross sections for third generation leptoquarks into the SS
topology as a function of $E_T^c$ are shown in Figs.\ \ref{ssjets}.  In Fig.\ 
\ref{ssjets}(a) one sees that the signal of $b$--$e$ leptoquarks is above the
background for $E_T^c \lesssim 200$ GeV while $b$--$\tau$ leptoquark signal is
above the background for $200$ GeV $\lesssim E_T^c \lesssim 430$ GeV.  The
$b$--$\tau$ leptoquark can probably only be distinguished from the mSUGRA case
6 if one demands $E_T^c > 400$ GeV.  In Fig.\ \ref{ssjets}(b) $t$--$e$ and
$t$--$\tau$ leptoquarks with a mass of 500 GeV are shown to be well above the
background and to lay between the mSUGRA cases.

\begin{figure}
\parbox[c]{6.0in}{
\mbox{\qquad\epsfig{file=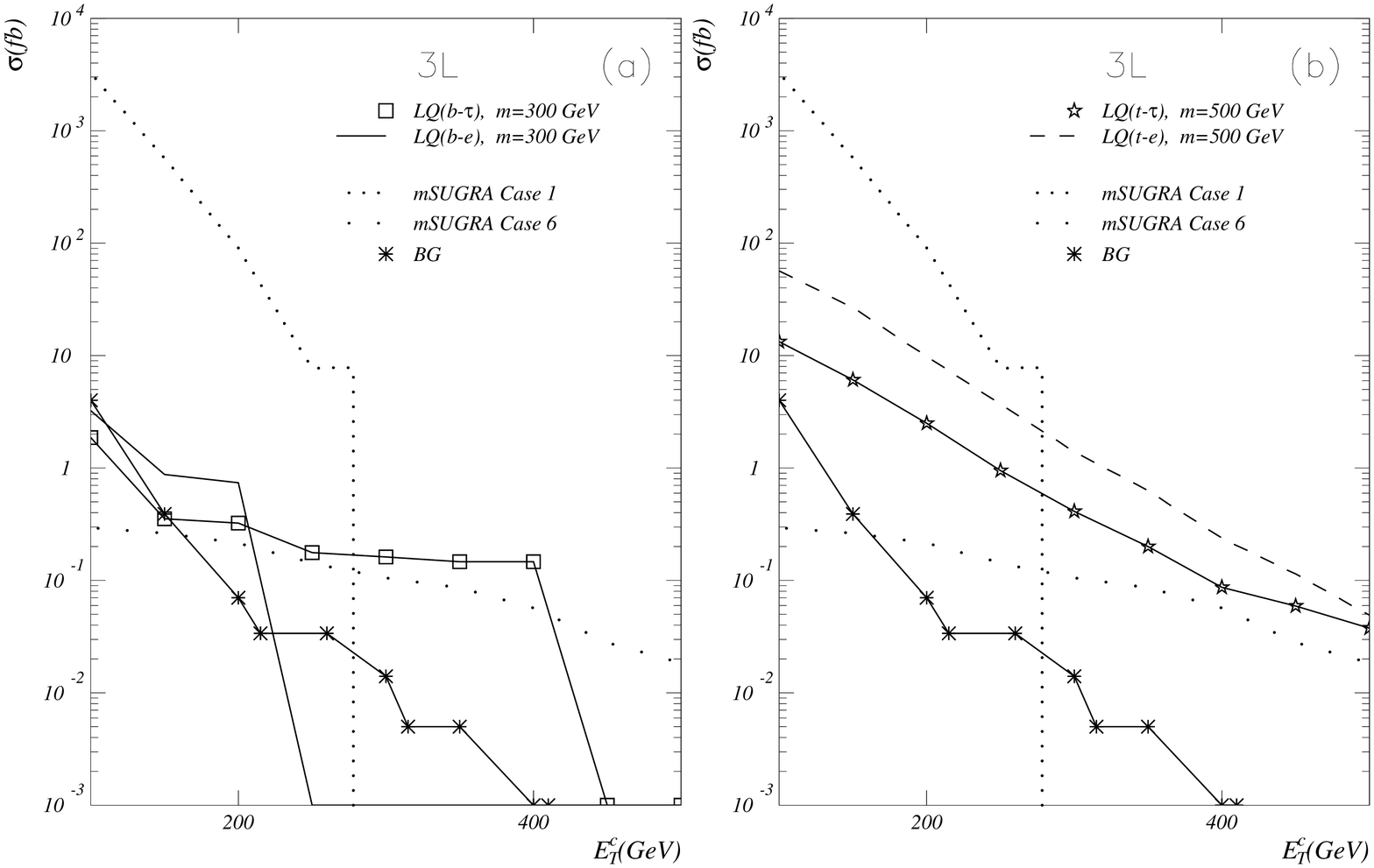,width=\linewidth}}
  }
\caption{Same as Fig.\ \protect\ref{1ljets} for 3L events.}
\label{3ljets}
\end{figure}

Finally in Figs.\ \ref{3ljets} the behavior of the cross sections for the 3L
topology as a function of $E_T^c$ is presented for the scalar leptoquarks as
well as for the mSUGRA cases and SM backgrounds.  The cross sections for
$b$--$e$ leptoquarks of 300 GeV shown in Fig.\ \ref{3ljets}(a) are above the
background only for $E_T^c \lesssim 200$ GeV.  The the production cross
section of $b$--$\tau$ leptoquarks is always above the background and it
presents a flat plateau in the region where $100$ GeV $ \lesssim E_T^c
\lesssim 400$ GeV.  In Fig.\ \ref{3ljets}(b) we see again that $t$--$e$ and
$t$--$\tau$ leptoquarks of 500 GeV are above the background and right in
between the mSUGRA extreme cases.

In our analyses we observed that third generation leptoquark cross sections
are generally above the SM background in all multilepton topologies we have
investigated. Moreover, the leptoquarks signals are of the same magnitude of
mSUGRA cross sections, making it rather difficult to distinguish SUSY events
from leptoquark ones.  It is clear that one has to investigate more carefully
the possibility of mistaken third generation leptoquarks for SUSY in the
multilepton channels.  Observation of the signal in several multilepton
channels is crucial to try to identify the source of new physics but this may
turn out to be a great challenger.


\section{Conclusion}

In this work, we analyzed the multilepton signals for third generation
leptoquarks. We showed that the analyses designed to discover gluinos and
squarks via multilepton events are also rather good to select third generation
leptoquarks. We concluded that for third generation leptoquarks with masses of
several hundred GeV, the leptoquark signal is not only above the standard
model backgrounds, but also of the same order of the expected mSUGRA cross
sections. Therefore, the observation of an excess of multilepton events
accompanied by jets and missing $E_T$ can be due to leptoquarks or
supersymmetric particles. Since the leptoquark mass reconstruction is usually
not efficient, due to the presence of neutrinos in many decays, there is no
clear footprint of leptoquarks in this class of events. Therefore, the origin
of the multilepton events can only the establish looking at other topologies,
for instance, multilepton events without the presence of jets, which are
characteristic of $\chi^2_0$--$\chi^\pm$ production in some regions of the
mSUGRA parameter space~\cite{xixi}.  It seems that unless nature is extremely
kind to us exhibiting signals of new physics in many different channels, an
observed signal in any of the four discussed channels at the LHC cannot be
uniquely interpreted as due to the production of SUSY particles.  Even if
observation is accomplished in all four channels analyzed in this work, it may
still not be possible to distinguish between leptoquarks and supersymmetric
particles.


\section{Acknowledgments}

We would like X.\ Tata for useful discussions.  This work was
supported by Conselho Nacional de Desenvolvimento Cient\'{\i}fico e
Tecnol\'ogico (CNPq), by Funda\c{c}\~ao de Amparo \`a Pesquisa do
Estado de S\~ao Paulo (FAPESP), and by Programa de Apoio a N\'ucleos
de Excel\^encia (PRONEX).



\begin{thebibliography}{99}



\bibitem{comp} For a review see, W.\ Buchm\"uller, Acta Phys.\ Austr.\ 
  Suppl.\ {\bf XXVII}, 517 (1985).

\bibitem{af} L.\ Abbott and E.\ Farhi, Nucl.\ Phys.\ {\bf B189}, 547 (1981).
  
\bibitem{tec} S.\ Dimopoulos, Nucl.\ Phys.\ {\bf B168}, 69 (1981); E.\ 
  Farhi and L.\ Susskind, Phys.\ Rev.\ D{\bf 20}, 3404 (1979); J.\ 
  Ellis {\em et.\ al.\/}, Nucl.\ Phys.\ {\bf B182}, 529 (1981).
  
\bibitem{gut} See, for instance, P.\ Langacker, Phys.\ Rep.\ {\bf 72},
  185 (1981); J.\ L.\ Hewett and T.\ G.\ Rizzo, Phys.\ Rep.\ {\bf
  183}, 193 (1989).

  
\bibitem{paige}H.\ Baer, C.\ Chen , F.\ Paige, and X.\ Tata, 
Phys. \ Rev.\ {\bf D53}, 6241 (1996). 


\bibitem{fut:pp} O.\ J.\ P.\ \'Eboli and A.\ V.\ Olinto, Phys.\ Rev.\ D{\bf
    38}, 3461 (1988); J.\ L.\ Hewett and S.\ Pakvasa, {\em ibid.}
    {\bf 37}, 3165 (1988); J.\ Ohnemus, S.\ Rudaz, T.\ F.\ Walsh, and
    P.\ Zerwas, Phys.\ Lett.\ {\bf B334}, 203 (1994); O.\ J.\ P.\
    \'Eboli and J.\ E.\ Cieza Montalvo, Phys.\ Rev.\ D{\bf 50}, 331
    (1994).  J.\ Bl\"umlein, E.\ Boos, and A.\ Krykov, hep-ph/9610408;
    M.\ Kr\"amer, {\em et al.}, Phys.\ Rev.\ Lett.\ {\bf 79}, 341
    (1997); J.\ L.\ Hewett and T.\ Rizzo, hep-ph/9703337; T.\ Rizzo,
    hep-ph/9609267; M.\ S.\ Berger and W.\ Merritt, hep-ph/9611386;
    O.\ J.\ P.\ \'Eboli, R.\ Z.\ Funchal, and T.\ L.\ Lungov, Phys.\
    Rev.\ D{\bf 57}, 1715 (1998); J.\ E.\ Cieza Montalvo {\em et al.},
    hep-ph/9805472, to appear in Phys.\ Rev.\ D.

    
  \bibitem{teva} CDF Collaboration, F. Abe {\em et. al.}, Phys. Rev. Lett.
    {\bf 78}, 2906 (1997).


\bibitem{shanker} O.\ Shanker, Nucl.\ Phys.\ {\bf B204}, 375 (1982).
  
\bibitem{fcnc} W.\ Buchm\"uller and D.\ Wyler, Phys.\ Lett.\ {\bf
    B177}, 377 (1986); J.\ C.\ Pati and A.\ Salam, Phys.\ Rev.\ D{\bf
    10}, 275 (1974).
  
\bibitem{gbjkm} G.\ Bhattacharyya, J.\ Ellis, and K.\ Sridhar, Phys.\ 
  Lett.\ {\bf B336}, 100 (1994); erratum {\em ibid.\/} {\bf B338}, 522
  (1994); O.\ J.\ P.\ \'Eboli, M.\ C.\ Gonzalez-Garcia, and J.\ K.\ 
  Mizukoshi, Nucl.\ Phys.\ {\bf B443}, 20 (1995); Phys.\ Lett.\ {\bf
    B396}, 238 (1997).
  
\bibitem{leurer}M.\ Leurer, Phys.\ Rev.\ Lett.\ {\bf 71}, 1324 (1993);
  Phys.\ Rev.\ D{\bf 49}, 333 (1994); S.\ Davidson, D.\ Bailey, and
  A.\ Campbell, Z.\ Phys.\ {\bf C61}, 613 (1994).


\bibitem{buch} W.\ Buchm\"uller, R.\ R\"uckl, and D.\ Wyler, Phys.\ 
  Lett.\ {\bf B191}, 442 (1987).


\bibitem{pythia} T.\ Sj\"ostrand, Computer Phys.\ Commun.\ {\bf 82}, 74
  (1994).

  
\bibitem{xixi} H.\ Baer, C.\ Chen, F.\ Paige, and X.\ Tata, Phys. Rev. {\bf
    D50}, 4508 (1994).


\end{thebibliography}
\end{document}